\documentclass[seceq]{ptptex}

\usepackage{graphicx}
\usepackage{wrapft}

\def\gtsim{\mathrel{\hbox{\raise0.2ex
   \hbox{$>$}\kern-0.75em\raise-0.9ex\hbox{$\sim$}}}}
\def\ltsim{\mathrel{\hbox{\raise0.2ex
   \hbox{$<$}\kern-0.75em\raise-0.9ex\hbox{$\sim$}}}}
\newcommand{\bm}[1]{\mbox{\boldmath $#1$}}

\title{
Efficient Method for Quantum Number Projection and Its Application
to Tetrahedral Nuclear States
}

\author{
Shingo \textsc{Tagami}$^1$, Yoshifumi R. \textsc{Shimizu}$^1$,
and Jerzy \textsc{Dudek}$^2$
}

\inst{
$^1$Department of Physics, Faculty of Sciences, Kyushu University,
Fukuoka 812-8581, Japan\\
$^2$Institut de Recherches Subatomiques, IN$_2$P$_3$-CNRS/Universit\'e
Louis Pasteur, F-67037 Strasbourg, France
}

\abst{
We have developed an efficient method for quantum number projection
from most general HFB type mean-field states, where all the symmetries
like axial symmetry, number conservation, parity and time-reversal
invariance are broken. 
Applying the method, we have microscopically calculated,
for the first time, the energy spectra based on the exotic
tetrahedral deformation in $^{108,110}$Zr.
The nice low-lying rotational spectra, which have all characteristic features
of the molecular tetrahedral rotor, are obtained for large tetrahedral
deformation, $\alpha_{32} \gtsim 0.25$, while the spectra are of
transitional nature between vibrational and rotational with rather
high excitation energies for $\alpha_{32}\approx 0.1-0.2$.
}

\begin{document}

\maketitle

\section{Introduction}

The progress of radioactive beam facilities in these years provides us
a great possibility to explore nuclear regions with various combinations
of neutron and proton numbers, where many interesting new phenomena
have been predicted.  In the present work, we would like to focus,
among others,
on the exotic shape with high rank point group symmetry~\cite{DGM10},
i.e., the tetrahedral nuclear states.
This ``pyramid-like'' shape leads to the extra stability in shell energy
due to the higher symmetry,
e.g., the appearance of fourfold degenerate orbits.
It has been suggested that such states appear as low-lying states,
or even as a ground state, in some nuclei
around the ``tetrahedral-closed shell'' nuclei;~\cite{Dudek02}
see also Refs.~\citen{OS71,LD94,TYM98,YMM01}.
In these earlier studies, the mean-field approaches are employed
to search for minimum configurations with the tetrahedral shape.
However, the quantum excitation spectra
and the properties of electromagnetic transition rates between them
are necessary to definitely identify the tetrahedral states.
The quantum number projection,
especially the angular momentum projection,
from mean-field states is useful for such a purpose.

Note that the tetrahedral shape is neither axially symmetric
nor reflection symmetric;
the number conservation is lost if the pairing correlation is included,
and the time-reversal invariance is broken if the cranking procedure
is used to investigate the collective rotational bands.
Therefore, the most general HFB type mean-field states without symmetry
restrictions are necessary to describe the tetrahedral nuclei.
Recently we have developed an efficient method for quantum number projection
from such most general HFB type states.~\cite{TS12}
Employing this method, in this work, we present the quantum spectra
of the nuclear tetrahedral rotor, which are obtained microscopically
for the first time, and discuss briefly its characteristic properties.
More detailed investigations will be reported elsewhere.

\section{Efficient Method for Projection and GCM}

Since the detailed information is published in Ref.~\citen{TS12},
here we briefly discuss the essential points of our method.
The projector is of the form
$\hat P_\alpha = \int g_\alpha(\bm{x}){\hat D}(\bm{x})d\bm{x}$,
where ${\hat D}(\bm{x})$ is a unitary transformation of symmetry operations,
e.g., the rotation, and the main task is to calculate a quantity
$\langle \Phi | {\hat O} {\hat D}(\bm{x})| \Phi' \rangle$
for an arbitrary operator $\hat O$ and HFB type states
$|\Phi \rangle$ and $|\Phi' \rangle$
at the mesh points of integration over the parameter space $(\bm{x})$.
The general HFB state is defined by the quasiparticle operators,
\begin{equation}
 \hat \beta_{k}^{}| \Phi \rangle =0 \ (k=1,2,...,M),\qquad
 \hat \beta_{k}^\dagger = \sum_{l} \left[
  U_{lk}\hat c_{l}^\dagger + V_{lk}\hat c_{l}^{}
 \right],
\label{eq:GBtra}
\end{equation}
where $(\hat c_{l}^\dagger,\hat c_{l}^{})$ are the creation and annihilation
operators of the states within the original basis,
for which the unitary transformation $\hat D$ is represented by
a matrix $D=(D_{l'l})$,
\begin{equation}
 \hat D \hat c_l^\dagger \hat D^\dagger
 = \sum_{l'} D_{l'l} \hat c_{l'}^\dagger.
\label{eq:Dgmat}
\end{equation}
Then, the quantity
$\langle \Phi | {\hat O} {\hat D}(\bm{x})| \Phi' \rangle$
can be expressed in terms of the matrix elements of $\hat O$,
the transformation matrix $D$, $(U,V)$ amplitudes in Eq.~(\ref{eq:GBtra})
and the similar amplitudes $(U',V')$ for $|\Phi' \rangle$.
All the matrices have sizes $M \times M$ and each matrix manipulation
requires $O(M^3)$ operations.
This is the difficulty for the projection from the HFB state
with a large model space; for example,
if we take $N^{\rm max}_{\rm osc}=20$ spherical harmonic oscillator shells,
the dimension becomes $M=3,542$.
If the pairing correlation is neglected and the calculation is restricted
to the HF state, it is shown that the matrix operations reduces to
$O(MN^2)$ for one-body operators,
where $N$ is the number of particles and is at most 200 or so.

The method we have taken to circumvent this difficulty can be summarized
by the following two points:
\begin{itemize}

\item Basis truncation in terms of the canonical basis;
with
\begin{equation}
 \hat b_k^\dagger = \sum_l W_{lk} \hat c_l^\dagger , \quad
 \langle \Phi|\hat b_{k}^\dagger \hat b_{k}^{}|\Phi\rangle
 =\delta_{kk'}v^2_k \quad (v^2_k\mbox{ in descending order}),
\label{eq:Wcanb}
\end{equation}
the $P$ space is defined as $\{k=1,2,...,L_p(\epsilon);\, v_k^2 \ge\epsilon\}$
with a small number $\epsilon$.

\item Full use of the Thouless form with respect to a Slater determinant;
\begin{equation}
|\Phi\rangle = n \exp{
\biggl(\sum_{l>l'}Z_{ll'}a^\dagger_la^\dagger_{l'}\biggr)}
 |\phi\rangle,\quad
 |\phi\rangle\equiv \prod_{k=1}^N b^\dagger_k|0\rangle,\quad
 a^\dagger_k=\left\{\begin{array}{ll}
  b^{}_k & k \le N \cr b^\dagger_k & k > N \end{array}\right..
\label{eq:ThZSD}
\end{equation}

\end{itemize}
In order to avoid the sign problem
of norm overlap for general HFB states~\cite{Rob09},
especially those without the time-reversal invariance, one has to calculate
the pfaffian with Thouless amplitudes.
Then it is easy to show that the matrix operations for one-body operators can
be reduced to $O(ML^2_p(\epsilon))$ by the canonical basis truncation above.
However, the Thouless amplitude with respect to nucleon vacuum diverges
when the  pairing correlation is vanishing or the HFB state is orthogonal
to the nucleon vacuum, e.g., a quasiparticle excited state.
The Thouless form with respect to a Slater determinant in Eq.~(\ref{eq:ThZSD})
solves this difficulty because then the amplitude
$Z \sim u_k/v_k$ for $k \le N$ and $Z \sim v_k/u_k$ for $k > N$
in the diagonal representation, and the truncation scheme can be
utilized for any HFB type mean-field states.
In addition, the calculation can be further reduced because of
the simple treatment of the core space (deep hole states) defined as
$\{k=1,2,...,L_o(\epsilon);\, u_k^2=1-v^2_k \ge\epsilon\}$.

\begin{wrapfigure}{r}{6.7cm}
\vspace*{-1mm}
\begin{center}
\includegraphics[width=0.46\textwidth]{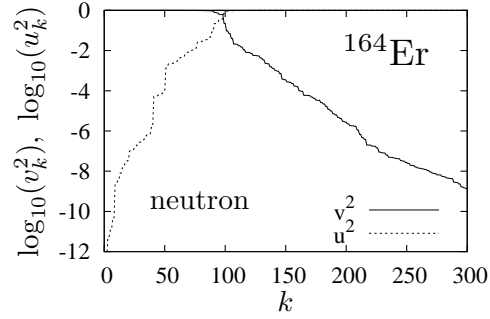}
\vspace*{-3mm}
\caption{
Probabilities $v_k^2$ and $u_k^2$
as functions of the number $k$ of the canonical basis.
}
\label{fig:Ervvuu}
\end{center}
\vspace*{-1mm}
\end{wrapfigure}

\begin{figure}[!ht]
\vspace*{-1mm}
\begin{center}
\hspace*{2mm}
\includegraphics[width=0.85\textwidth]{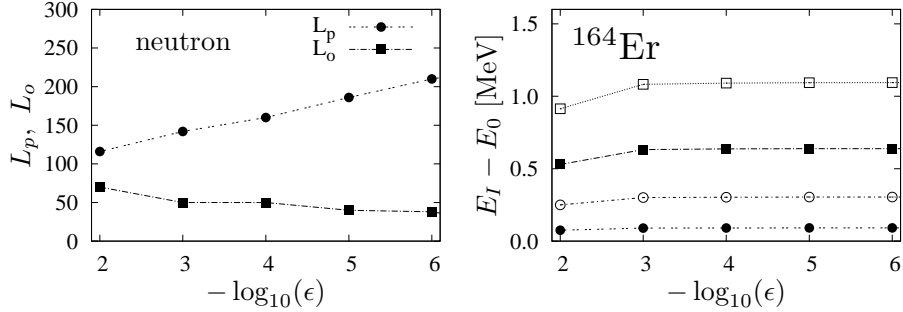}
\vspace*{-4mm}
\caption{
Dimensions of the $P$-space $L_p$ and the core space $L_o$
as functions of the model space truncation parameter $\epsilon$ (left),
and the rotational spectra ($I=2,..,8$)
as functions of the truncation parameter $\epsilon$ for $^{164}$Er (right).
}
\label{fig:Erconv}
\end{center}
\vspace*{-1mm}
\end{figure}

Examples of the occupation and empty probabilities,
$v^2_k$ and  $u^2_k$, are shown in Fig.~\ref{fig:Ervvuu},
and the resultant two dimensions, $L_p(\epsilon)$ and $L_o(\epsilon)$,
as well as the calculated rotational spectra
are shown in Fig.~\ref{fig:Erconv}.
It can be seen that the truncation parameter
$\epsilon\approx 10^{-3}- 10^{-4}$ is sufficient,
and then the necessary $P$ space dimension is $L_p \approx 150$
($L_o\approx 50$),
which is much smaller than the original basis size $M\approx 3,000$.
Thus the calculational effort is dramatically reduced.

As for the choice of the Hamiltonian,
it may be desirable to employ effective interactions
like Skyrme and Gogny forces, but they have problems
for the projection and the GCM calculations due to their density-dependence.
Thus, for the illustration of our method,
we use the following schematic multi-separable interaction:
\begin{equation}
\hat H = \hat h -\frac{1}{2} \chi \sum_\lambda
:\hat F^\dagger_\lambda\cdot\hat F^{}_\lambda:
-\sum_{\lambda,\tau={\rm n,p}}g^\tau_\lambda\,
\hat G^{\tau\dagger}_\lambda\cdot\hat G^\tau_\lambda,
\label{eq:Hamil}
\end{equation}
where the single-particle part $\hat h$ is composed of
the Woods-Saxon potential.
The two-body interaction is constructed from
the isoscalar particle-hole channel operators
$F^{}_{\lambda\mu}=\sum_{\tau={\rm n,p}}F^\tau_{\lambda\mu}$
and the pairing channel operators $G^\tau_{\lambda\mu}$ defined by
\begin{equation}
 F^\tau_{\lambda\mu}(\bm{r})=R^\tau_0\, \frac{dV^\tau_c(r)}{dr}\,
 Y_{\lambda\mu}(\theta,\phi),\quad
 G^\tau_{\lambda\mu}(\bm{r})=\left(\frac{r}{\bar R_0}\right)^\lambda
 \sqrt{\frac{4\pi}{2\lambda+1}}\, Y_{\lambda\mu}(\theta,\phi),
\label{eq:FG}
\end{equation}
with $V^\tau_c(r)$ and $R^\tau_0$ being the central part of
the Woods-Saxon potential and its radius parameters, respectively, and
$\bar R_0\equiv 1.2 A^{1/3}$ fm.  The multipolarities $\lambda=2,3,4$ 
are included in the p-h channel and $\lambda=0,2$ in the pairing channel.
As for the force strengths, we adopt the selfconsistent value for $\chi$,
while the values of $g^\tau_0$ are determined
according to even-odd mass differences ${\mit\Delta}_\tau$.
The ratio $g^\tau_2/g^\tau_0=13.6$ is chosen
to reproduce typical rotational bands in a rare-earth
and an actinide nucleus, see Ref.~\citen{TS12} for details.
It should be noted that this kind of schematic Hamiltonian
cannot be used to obtain the ground state energy and deformation;
it is only used to calculate the collective excitation spectra
based on a given mean-field state $|\Phi\rangle$.

\section{Spectra of Nuclear Tetrahedral Rotor}

The tetrahedra deformation can be conveniently described by
the $\alpha_{32}$ deformation without other terms
in the usual nuclear surface parametrization,
\begin{equation}
R(\theta,\varphi)=R_0 c_v(\{\alpha\})
\biggl(1+\sum_{\lambda\mu}\alpha^*_{\lambda\mu}Y_{\lambda\mu}(\theta,\varphi)
\biggr).
\label{eq:surf}
\end{equation}
Thus, it is clear that both the axial symmetry and parity are broken.
Now let us investigate what kinds of quantum spectra with definite spin-parity
are expected for the tetrahedrally deformed nuclear state.
For this purpose, it is interesting to consider the region near
one of double tetrahedral-closed shell nuclei,
$^{110}$Zr ($Z=40$ and $N=70$),
because experiments for neutron-rich Zr isotopes
have been done recently at RIKEN~\cite{Sumik11}.
In this experiment a new isomer with life time of about 600 nsec
has been found in a nucleus $^{108}$Zr, and speculated
as a possible low-lying tetrahedral state according to
a theoretical prediction~\cite{Schun04}.

\begin{figure}[!hbt]
\vspace*{-1mm}
\begin{center}
\hspace*{0mm}
\includegraphics[width=0.85\textwidth]{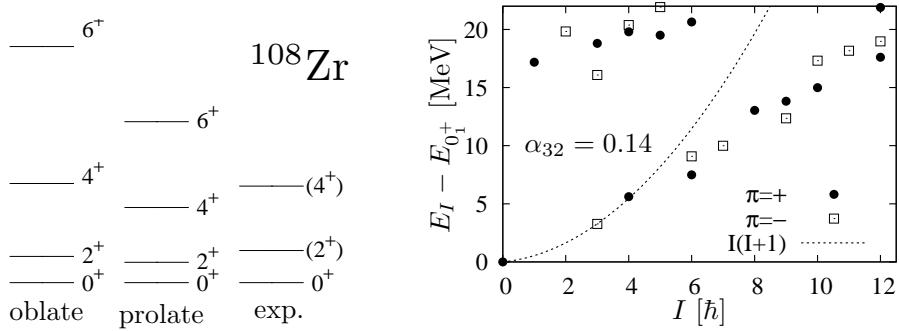}
\vspace*{-3mm}
\caption{
Calculated spectra of the ground state band in $^{108}$Zr (left),
and of tetrahedral states assuming $\alpha_{32}=0.14$ (right).
Dotted curves denote the $I(I+1)$ spectra going through $3^-_1$.
Although higher excited states may not have realistic meaning,
they are included to show what kind of spin-parity appears.
}
\label{fig:Zr108}
\end{center}
\vspace*{-2mm}
\end{figure}

We apply our quantum number projection method to $^{108,110}$Zr nuclei.
In order to obtain a reliable mean-field for the ground states,
we have performed the paired Woods-Saxon Strutinsky calculation
of Ref.~\citen{TST10} (with the Wyss-2 potential parameter set).
Since the pairing model space is cut off in these calculations,
we do not include the cut-off factor,
which was introduced in Ref.~\citen{TS12}
for the pairing operators $G^\tau_{\lambda\mu}$ in Eq.~(\ref{eq:Hamil}),
in the present projection calculation.
The pairing force strength $g^\tau_0$ are fixed according to those results
with the same ratio $g^\tau_2/g^\tau_0=13.6$ as previously.
The model space size $N_{\rm osc}^{\rm max}=18$ and
the truncation parameter $\epsilon=10^{-4}$ are used.
Because the cranking procedure is important to evaluate the moment of
inertia~\cite{TS12},
we use a small cranking frequency $\hbar\omega_{\rm rot}=0.01$ MeV
about the $x$-axis in all calculations.

We compare calculated ground state bands
in $^{108}$Zr with experimental data~\cite{Sumik11} in Fig.~\ref{fig:Zr108}.
Since the total energies of prolate and oblate minima
are almost the same, we show spectra on both of them.
The oblate spectra better agrees with data,
though the prolate ground state energy is slightly lower.
In any case, this confirms that our Hamiltonian is a reasonable one.
The calculated spectra of tetrahedral states
with assumed deformation~\cite{Schun04} $\alpha_{32}=0.14$
are also shown in Fig.~\ref{fig:Zr108}.
Although the spectra are not so rotational-like,
their spin-parity appears to be characteristic to the simplest
singlet representation of the tetrahedral rotor~\cite{Herzberg}
corresponding to the closed shell, i.e.,
$0^+$, $3^-$, $4^+$, $6^{\pm}$ (doublet), etc.,
even for a neutron non-closed nucleus $^{108}$Zr.
This is because the totally symmetric state is realized
owing to the pairing correlation.
If the pairing is switched off,
two neutrons partially fill the last fourfold generate orbit
and the resultant Slater determinant is not tetrahedrally symmetric.
Then, the spectra are more complex combinations
of various irreducible representations.

\begin{figure}[!ht]
\vspace*{-1mm}
\begin{center}
\hspace*{0mm}
\includegraphics[width=0.85\textwidth]{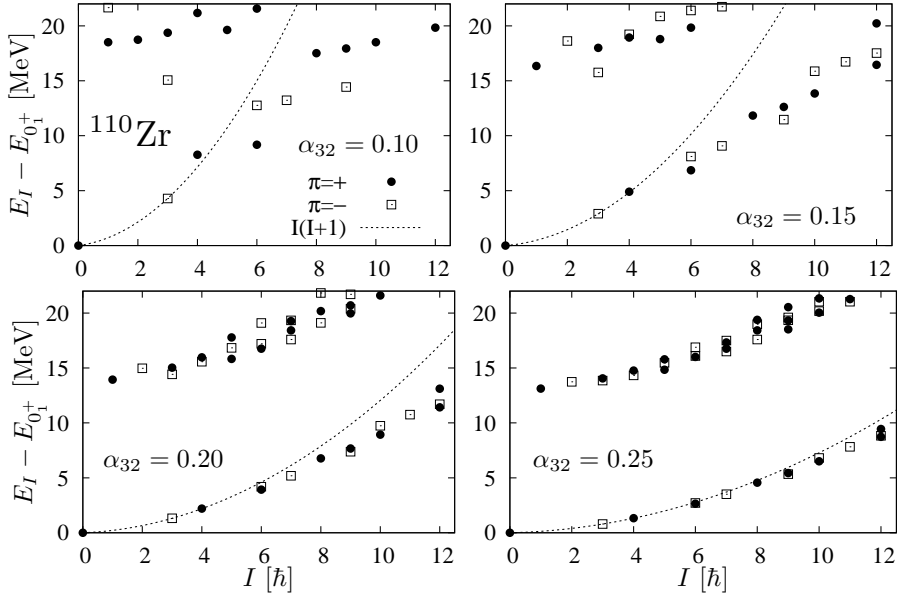}
\vspace*{-3mm}
\caption{
Calculated spectra of tetrahedral states in $^{110}$Zr with
various values of $\alpha_{32}$.
These results correspond to projection from {\em a single HFB wave function};
the high lying bands serve merely the illustration of the symmetry properties
and are not meant to be compared with experiment.
}
\label{fig:Zr110}
\end{center}
\vspace*{-1mm}
\end{figure}

\begin{figure}[t]
\begin{center}
\hspace*{0mm}
\includegraphics[width=0.8\textwidth]{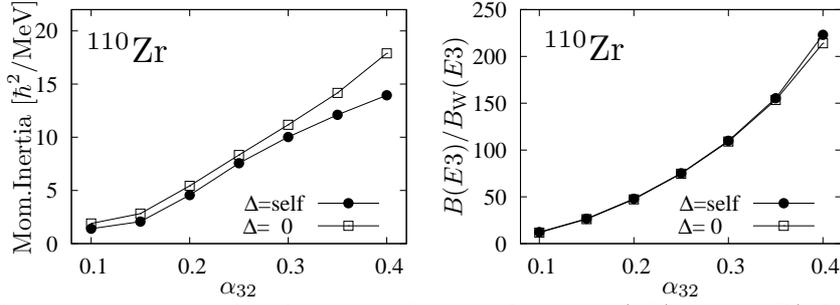}
\vspace*{-4mm}
\caption{
Calculated moment of inertia deduced from the $3^-_1$ energy (left)
and the $B(E3:3^-_1 \rightarrow 0^+_1)$ in Weisskopf unit (right)
as functions of $\alpha_{32}$ for $^{110}$Zr.
The rigid-body inertia is 35.1 [$\hbar^2$/MeV].
}
\label{fig:Zr110MB}
\end{center}
\vspace*{-2mm}
\end{figure}

In Fig.~\ref{fig:Zr110} the calculated tetrahedral spectra
of the double closed nucleus $^{110}$Zr,
for which the mean-field state is tetrahedrally symmetric and gives
the same characteristic spectra even in the unpaired case,
are depicted for various values of $\alpha_{32}$.
It is clearly seen that the spectra gradually change to rotational ones
at considerable deformation $\alpha_{32}=0.25$,
while they are more like vibrational or transitional
in $0.1 \ltsim \alpha_{32} \ltsim 0.20$;
the $3^-_1$ energy quickly decreases as $\alpha_{32}$ increases.
This kind of phase transitions is well-known for the quadrupole deformation.
We plot the moment of inertia estimated by $6/E(3^-_1)$
in Fig.~\ref{fig:Zr110MB}, where the results with switching off
the pairing correlation are also included.
Note that the effect of the pairing correlation is not so strong.
This may be partly because this nucleus is double tetrahedral-closed
and the pairing gaps of neutron and proton are not so large.
Compared to the rigid body value,
$\frac{2}{5}AM{\bar R}_0^2=35.1$ $\hbar^2$/MeV,
the moment of inertia is rather small
even at $\alpha_{32}\approx 0.4$ without pairing correlation.
The calculated $B(E3:3^-_1 \rightarrow 0^+_1)$ value
is also shown in Fig.~\ref{fig:Zr110MB}, which well corresponds
to the rotor model estimate,
$(2-\delta_{K0})
|\langle \Phi|r^3Y_{3K}|\Phi\rangle|^2 \langle 3 K 3 -K| 0 0\rangle^2$,
with $K=2$.  Again, they become very large,
more than 100 Weisskopf unit, for $\alpha_{32}\gtsim 0.3$.

In summary, we have calculated the spectra of the nuclear tetrahedral
rotor for the first time by employing the quantum number projection method.
By increasing the tetrahedral deformation, the spectra gradually change
from vibrational to rotational just like in the spherical-to-deformed
phase transition in the case of quadrupole deformation.
It is quite interesting to measure the characteristic spectra
of the tetrahedral rotor,
which is a clear indication of the exotic tetrahedral deformation
in atomic nuclei.

\section*{Acknowledgements}
This work is supported
by Grant-in-Aid for Scientific Research~(C)
No.~22540285 from Japan Society for the Promotion of Science.

\end{document}